%% file: main.tex
\documentclass[twocolumn,superscriptaddress,amsmath,amssymb,aps,pra,floatfix,longbibliography]{revtex4-2}

\usepackage{graphicx}

\usepackage[utf8]{inputenc}
\usepackage[T1]{fontenc}

\begin{document}

\title{Interaction-induced period shift of the Kohn oscillation in a parity-time-symmetric harmonic trap}
\author{Gaoqing Meng}
\affiliation{
Hebei Key Laboratory of Physics and Energy Technology, Department of Mathematics and Physics, North China Electric Power University, Baoding, Hebei 071003, China
}
\author{Mingshu Zhao}
\email{zmshum@ncepu.edu.cn}
\affiliation{
Hebei Key Laboratory of Physics and Energy Technology, Department of Mathematics and Physics, North China Electric Power University, Baoding, Hebei 071003, China
}

\begin{abstract}
The center of mass of a Bose--Einstein condensate in a harmonic trap oscillates at the trap frequency independently of the interactions.
Without interactions a parity-time-symmetric gain-loss gradient leaves this frequency unchanged, and a cloud of the ground-state width is translated rigidly while the atom number is periodically modulated.
With interactions the protection is lost, because the gain and the loss change the number of atoms; the modulated atom number drives the width of the cloud, and the width acts back on the center of mass.
At small gain the period grows quadratically with the gain strength, with a coefficient that in the Thomas--Fermi regime is set by the size of the cloud alone.
The family of periodic orbits continued from the Kohn oscillation can be followed up to a critical gain, where it comes into resonance with highly excited states of the trap, and over the computed families this critical gain decreases with the amplitude of the oscillation and with the size of the cloud.
A five-variable moment model reproduces the period and the width of the cloud along the orbits nearly up to the critical gain.
\end{abstract}

\maketitle

\section{Introduction}

In a harmonic trap the oscillation of the center of mass (COM) is a collective mode whose frequency is unaffected by the interactions and equals the trap frequency.
Kohn proved the corresponding statement for the cyclotron motion of interacting electrons~\cite{kohn1961cyclotron}, and the same argument covers electrons in parabolic quantum wells~\cite{brey1989optical} and, as the harmonic-potential theorem, any interacting system in a harmonic potential~\cite{dobson1994harmonic}, atoms in magnetic and optical traps included.
The proof requires two conditions: the trap potential must be exactly quadratic, and the interaction must depend only on relative coordinates.
For a condensate described by the Gross--Pitaevskii (GP) equation the theorem applies to the condensate wave function itself.
Any solution can be set into rigid oscillation in the trap, so that every stationary state generates a periodic orbit whose density is translated rigidly~\cite{garciaripoll2001construction,moulieras2012wave}.
Because its frequency is set by the trap alone, the dipole mode serves as the standard calibration of trap frequencies in cold-atom experiments~\cite{dalfovo1999theory}.

The conditions of the theorem have been relaxed in several ways in previous work.
Anharmonic corrections to the trap couple the COM to the internal motion and generate harmonics of the dipole frequency~\cite{ott2003nonlinear}.
A scattering length made position dependent by the inhomogeneous field near a Feshbach resonance~\cite{aljibbouri2013breakdown}, and the retarded, nonlocal thermo-optic interaction in a photon condensate~\cite{stein2019collective}, are predicted to lower the dipole frequency.
Spin-orbit coupling ties the COM to the spin and makes the dipole oscillation anharmonic~\cite{zhang2012collective}, and in a two-component condensate only the total COM is protected, the spin-dipole frequency softening as the mixture approaches immiscibility~\cite{sartori2015spin,bienaime2016spin}.

In the system studied here the trap remains harmonic and the interaction local, but the potential acquires an imaginary part linear in position, so that atoms are added on one side of the trap and removed on the other, at rates that grow with the distance from the center.
Such a potential is parity-time (PT) symmetric.
Potentials of this kind can have real spectra without being Hermitian~\cite{bender1998real,elganainy2018nonhermitian} and have been realized with light in coupled waveguides~\cite{guo2009observation,ruter2010observation} and in warm rubidium vapor~\cite{zhang2016observation}.
The harmonic trap with a linear gain-loss term is special among them.
A similarity transformation maps it onto the Hermitian oscillator at every gain strength, so that its spectrum is real with no symmetry-breaking threshold, and without interactions every wave packet recurs after one trap period, a packet of the ground-state width being translated rigidly while the atom number is modulated~\cite{longhi2016pt}.

The gain and the loss, however, change the number of atoms and with it the interaction energy, and this opens an indirect route by which the protection may fail.
The gain modulates the atom number, the modulation drives the width of the cloud, and the width acts back on the COM.
Whether the period is shifted in this way, by how much, and up to which gain a periodic oscillation persists are the questions addressed here.
The nonlinear problem in this potential has been studied for stationary modes, whose COM does not move, and for the spectra of their linearizations~\cite{zezyulin2012nonlinear,konotop2016nonlinear,chernyavsky2018krein}, including the ground state in the Thomas--Fermi limit, which exists only below a threshold gain~\cite{gallo2014thomas}.
In a harmonic trap with a localized gain-loss profile, pairs of stationary states merge and disappear as the gain grows, in a way reminiscent of PT symmetry breaking~\cite{achilleos2012solitons}.
For complex potentials whose imaginary part contains a quadratic term, which are not PT symmetric, the damping of the motion of linear and nonlinear waves in a harmonic trap has been studied~\cite{salerno2026dissipative}.
The linearization about the stationary ground state contains the dipole oscillation at infinitesimal amplitude.
At finite amplitude, however, the density crosses the trap and the atom number is modulated along the way, so that the objects to be computed are time-periodic solutions of large amplitude, continued from the Kohn oscillation to finite gain.

In this work we compute these orbits and their Floquet multipliers, which measure their linear stability.
At small gain the period grows quadratically with the gain strength, with a coefficient set by the size of the cloud.
The family of orbits continued from the Kohn oscillation can be followed up to a critical gain, where it comes into resonance with highly excited states of the trap, and orbits of smaller amplitude persist to larger gain.
A five-variable model built from the moment equations reproduces the period and the width of the cloud along these orbits nearly up to the critical gain.

\section{Model and linear limit}\label{sec:model}

\subsection{Gross--Pitaevskii equation with a gain-loss gradient}

\begin{figure}
\centering
\includegraphics{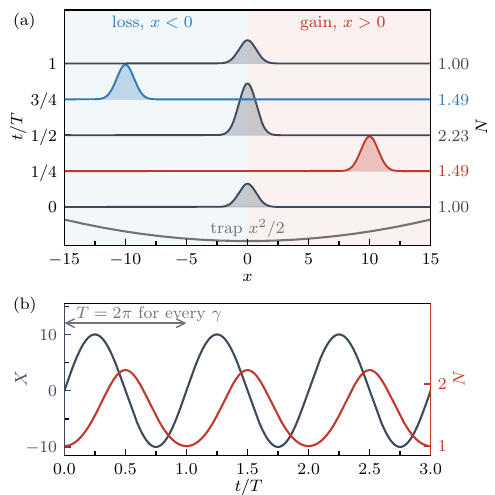}
\caption{The trap and its dynamics in the absence of interactions, from Eq.~\eqref{eq:linear} with $\gamma=0.02$ and $A=10$~\cite{longhi2016pt}.
(a) The density $|\psi|^2$ at five phases of one period, with $N$ on the right.
(b) $X$ and $N$ over three periods.}
\label{fig:linear}
\end{figure}

We consider the one-dimensional motion of a condensate in a harmonic trap of angular frequency $\omega$, to which an imaginary potential linear in position is added.
In the mean-field description the condensate wave function obeys
\begin{equation}
i\hbar\,\partial_t\psi=\Big[-\frac{\hbar^2}{2m}\partial_x^2+\frac12m\omega^2x^2+i\hbar\Gamma x+g_{\rm 1D}|\psi|^2\Big]\psi ,
\label{eq:gpe_dim}
\end{equation}
where $g_{\rm 1D}$ is the one-dimensional coupling constant; the arrangement is sketched in Fig.~\ref{fig:linear}(a).
The imaginary term makes the density grow at the local rate $2\Gamma x$, so that atoms are added where $x>0$ and removed where $x<0$, and the continuity equation carries a source, $\partial_t|\psi|^2=-\partial_xj+2\Gamma x|\psi|^2$ with $j$ the particle current.
Since $j$ vanishes at infinity, its divergence integrates to zero over the line, and the atom number $N=\int|\psi|^2dx$ obeys
\begin{equation}
\frac{dN}{dt}=2\Gamma XN ,
\label{eq:Ndot}
\end{equation}
where $X=\int x|\psi|^2dx/N$ is the position of the COM.
Equation~\eqref{eq:gpe_dim} thus describes an open system that exchanges atoms with reservoirs; for a condensate in a double well with balanced gain and loss, the corresponding many-particle master equation has been constructed, and its mean-field limit is a GP equation of this kind~\cite{dast2014quantum}.
A periodic solution requires $N(T)=N(0)$, that is, $\int_0^TX\,dt=0$, so that the gain and the loss cancel over each period while $N$ remains modulated within it.

Throughout, $t=0$ denotes an instant at which the COM crosses the trap center moving toward the gain side.
We measure length in units of $\sqrt{\hbar/m\omega}$, time in units of $1/\omega$, and the wave function in units such that $\int|\psi|^2dx=1$ at $t=0$; since the atom number is not conserved, this fixes the unit of $\psi$ and is not a normalization that holds at other times.
Equation~\eqref{eq:gpe_dim} becomes
\begin{equation}
i\partial_t\psi=\Big[-\tfrac12\partial_x^2+\tfrac12x^2+i\gamma x+g|\psi|^2\Big]\psi ,
\label{eq:gpe}
\end{equation}
with $\gamma=\Gamma\sqrt{\hbar/m\omega^3}$ and $g=g_{\rm 1D}N_0\sqrt{m}/(\hbar^{3/2}\omega^{1/2})$, where $N_0=N(t=0)$.
We use $g=100$ unless stated otherwise.
For this coupling the ground state of the trap at $\gamma=0$ is in the Thomas--Fermi (TF) regime.
Its chemical potential, $14.134$, lies within $0.2\%$ of the TF value $\mu=(3g/4\sqrt2)^{2/3}=14.116$, which gives the radius $R_{\rm TF}=\sqrt{2\mu}=5.313$, and its squared width is $\sigma_0^2=\langle x^2\rangle=5.696$.
The speed of sound at the center of the cloud, $\sqrt\mu$ in these units, is $c\approx1.1$~mm/s for $^{87}$Rb in a trap of $\omega=2\pi\times20$~Hz, comparable to typical experimental values~\cite{andrews1997propagation,zhao2025kolmogorov}; condensates of more atoms have larger $g$ and lie deeper in the TF regime.

\subsection{Symmetries}

At $\gamma=0$ Eq.~\eqref{eq:gpe} has the Kohn symmetry, the Galilean boost of the trapped problem.
If $\psi(x,t)$ is a solution, so is $\psi(x-X,t)\exp[i\dot Xx-\tfrac{i}{2}\int^t(\dot X^2-X^2)\,dt']$, where $X(t)$ is any trajectory satisfying $\ddot X=-X$.
Applied to a stationary state it generates a periodic orbit of period $2\pi$ and arbitrary amplitude, whose density is translated rigidly along that trajectory.

For $\gamma\neq0$ the potential $V=\tfrac12x^2+i\gamma x$ of Eq.~\eqref{eq:gpe} satisfies $V^*(-x)=V(x)$, which is the PT symmetry.
Reflection exchanges the region of gain with the region of loss, and time reversal converts gain into loss, so that only their combination leaves the dynamics invariant.
The equation is then reversible, in the sense that $\psi^*(-x,-t)$ is a solution whenever $\psi(x,t)$ is, and we write $\mathcal R\psi(x)=\psi^*(-x)$ for the reversing operation on states.
Because $\gamma\to-\gamma$ amounts to $x\to-x$, it suffices to consider $\gamma>0$.
A symmetric orbit, one mapped onto itself by $\mathcal R$, passes twice per period through a PT-symmetric state, $\psi(x)=\psi^*(-x)$ up to a global phase, at the instants when the COM crosses the trap center, which are $t=0$ and $t=T/2$ in Fig.~\ref{fig:linear}(a); the numerical method of Appendix~\ref{app:numerics} is built on this property.
Since $X$ is odd in time about each of these instants, such an orbit meets the condition $\int_0^TX\,dt=0$ by symmetry.

At a given $\gamma$ the periodic orbits are labeled by two parameters, the atom number and the amplitude of the oscillation, as the Kohn orbits are at $\gamma=0$.
We fix both at $t=0$: the normalization sets $N(0)=1$, and we prescribe the velocity of the COM, $\dot X(0)=p_0$.
At $\gamma=0$ the amplitude is then $A=p_0$; for $\gamma\neq0$ the amplitude $A$, defined as the largest $X$ along the orbit, follows from the computed orbit.
Followed in $\gamma$ at fixed $g$ and $p_0$, starting from the Kohn orbit at $\gamma=0$, the orbits form a family.
Unless stated otherwise we take $p_0=10$, so that the turning points, at $A\approx2R_{\rm TF}$, lie well outside the cloud.

\subsection{Linear limit}

The noninteracting limit of Eq.~\eqref{eq:gpe} is known in closed form~\cite{longhi2016pt}.
For $g=0$ the potential can be completed to a square, $\tfrac12x^2+i\gamma x=\tfrac12(x+i\gamma)^2+\tfrac12\gamma^2$, so the PT-symmetric oscillator is the Hermitian oscillator evaluated at the complex coordinate $x+i\gamma$.
The similarity transformation $e^{\gamma p}$, where $p=-i\partial_x$ is the momentum operator, translates the coordinate by $-i\gamma$ and carries it into the Hermitian oscillator, whose spectrum it reproduces displaced rigidly upward by $\tfrac12\gamma^2$.
The spectrum is therefore real for every $\gamma$ and remains equally spaced, so that every state recurs after the time $2\pi$ up to a global phase.
The reality of the spectrum is an instance of pseudo-Hermiticity~\cite{mostafazadeh2002pseudo}, and there is no threshold at which the symmetry breaks, in contrast to the potentials $x^2(ix)^\epsilon$ and to the two-level and double-well systems that are usually employed to introduce PT symmetry~\cite{bender1998real,bender2007making,elganainy2018nonhermitian,cartarius2012model}.

Continuing the coherent states of the Hermitian oscillator to the complex coordinate in the same way gives the wave-packet dynamics~\cite{longhi2016pt}.
The coherent state with $N=1$ at $t=0$ is a Gaussian of fixed width that follows the trap trajectory rigidly, $|\psi|^2=\pi^{-1/2}N(t)\,e^{-(x-X)^2}$, with
\begin{equation}
X=A\sin t ,\qquad N(t)=e^{2\gamma A(1-\cos t)} .
\label{eq:linear}
\end{equation}
The profile and its trajectory are thus those of the Hermitian trap, while $N$ varies by the factor $e^{4\gamma A}$ between its minimum and its maximum.
The two oscillations are a quarter period out of phase, since by Eq.~\eqref{eq:Ndot} it is the position that drives the atom number; as Fig.~\ref{fig:linear}(b) shows, $N$ is extremal at the center crossings and $X$ at the turning points.

\section{The period}\label{sec:ehrenfest}

The linear limit leaves the period at $2\pi$, so any shift must come from the interaction.

\subsection{Non-Hermitian Ehrenfest theorem}

For a wave function evolving under $i\partial_t\psi=(H_R+iW)\psi$ with $H_R$ and $W$ Hermitian, the expectation value $\langle O\rangle=\int\psi^*O\psi\,dx/N$ of a time-independent operator obeys~\cite{graefe2011wave,brody2012mixed}
\begin{equation}
\frac{d\langle O\rangle}{dt}=i\langle[H_R,O]\rangle+\langle\{W,O\}\rangle-2\langle W\rangle\langle O\rangle .
\label{eq:ehrenfest}
\end{equation}
The first term is the Ehrenfest theorem of the Hermitian problem; the second arises from the anti-Hermitian part $iW$ of the generator and the third from the time dependence of the normalization, $\dot N/N=2\langle W\rangle$.
For our problem $H_R=-\tfrac12\partial_x^2+\tfrac12x^2+g|\psi|^2$, the mean-field term being a real multiplication operator and therefore Hermitian at each instant, and $W=\gamma x$.

Taking $O=x$ and $O=p$ in turn, and writing $P=\langle p\rangle$, $S=\langle x^2\rangle-X^2$ for the squared width, and $C=\tfrac12\langle xp+px\rangle-XP$ for the position--momentum covariance, we find
\begin{equation}
\dot X=P+2\gamma S ,\qquad \dot P=-X+2\gamma C .
\label{eq:com}
\end{equation}
The interaction does not appear; its contribution to $\dot P$ vanishes because $\int\rho\,\partial_x\rho\,dx=0$ for $\rho=|\psi|^2$, exactly as in the Hermitian trap.
The gain enters instead through the last two terms of Eq.~\eqref{eq:ehrenfest}, and the resulting term $2\gamma S$ in $\dot X$ does not originate from a force.
The density grows on the gain side and decays on the loss side, so that the mean position drifts toward the gain side in the absence of any transport, at a rate proportional to $S$.
Eliminating $P$,
\begin{equation}
\ddot X=-X+2\gamma\big(C+\dot S\big).
\label{eq:Xdd}
\end{equation}

For a cloud whose shape does not change in time $\dot S=0$, and the covariance of a stationary profile vanishes, so that the COM obeys $\ddot X=-X$ for every $\gamma$; the rigid oscillation of the noninteracting limit is an instance of this.
A time-dependent $S$ is necessary for a shift of the period but not sufficient; $C+\dot S$ must also contain a component at the trap frequency in phase with $X$.
A noninteracting packet of another width breathes at twice the trap frequency and leaves the period unchanged; the interaction supplies such a component, because the width of an interacting cloud responds to its atom number, which is modulated along the orbit at the trap frequency.

\subsection{Moment equations and a five-variable model}

Applying Eq.~\eqref{eq:ehrenfest} to $x^2$ and to $xp+px$ and reducing to central moments gives
\begin{align}
\dot S&=2C+2\gamma M_3 ,\label{eq:Sdot}\\
\dot C&=\Pi-S+\tfrac12 gN\!\int\tilde\rho^2dx+2\gamma M_{21} ,\label{eq:Cdot}
\end{align}
where $\Pi=\langle p^2\rangle-P^2$ is the momentum variance and $\tilde\rho=\rho/N$ the normalized density, and where, with $\xi=x-X$ and $\eta=p-P$, $M_3=\langle\xi^3\rangle$ is the third central moment and $M_{21}=\langle\xi\eta\xi\rangle$ the symmetrized third central moment of position and momentum.
Appendix~\ref{app:moments} derives Eqs.~\eqref{eq:Sdot} and \eqref{eq:Cdot}.
In Eq.~\eqref{eq:com} the gain multiplies the second moments $S$ and $C$, whereas in Eqs.~\eqref{eq:Sdot} and \eqref{eq:Cdot} it multiplies only the third moments $M_3$ and $M_{21}$, which measure the asymmetry of the cloud about its center.

To close the equations we assume, as in the time-dependent variational analysis of a trapped condensate~\cite{perezgarcia1996low} and in the quadratic-phase approximation of the method of moments~\cite{perezgarcia2007method}, that the cloud keeps a fixed shape while its width changes, $\tilde\rho=S^{-1/2}F(\xi/\sqrt S)$ with $\int F\,du=\int u^2F\,du=1$, and that its phase is quadratic in $\xi$ with a curvature set by $C$.
Two integrals of that profile enter, $\kappa=\int F^2du$ and $q=\int[(\sqrt F)']^2du$, the second being its quantum pressure, and with them the two moments that Eqs.~\eqref{eq:Sdot} and \eqref{eq:Cdot} leave open become
\begin{equation}
\int\tilde\rho^2dx=\frac{\kappa}{\sqrt S},\qquad \Pi=\frac{q}{S}+\frac{C^2}{S}.
\label{eq:closure}
\end{equation}
A symmetric $F$ has $M_3=M_{21}=0$, so that $\gamma$ disappears from Eqs.~\eqref{eq:Sdot} and \eqref{eq:Cdot}, leaving the factor $N(t)$ of the interaction term as the only route by which the gain reaches $S$.

For $q$ we take $1/4$, the value for a Gaussian $F$ as in the noninteracting cloud, while $\kappa$ is fixed by requiring the correct squared width at $\gamma=0$.
With $C=0$ and $N=1$, Eqs.~\eqref{eq:Cdot} and \eqref{eq:closure} give $q/\sigma_0^2-\sigma_0^2+g\kappa/(2\sigma_0)=0$, with $\sigma_0^2$ taken from the numerical ground state.
For $g=100$ this gives $\kappa=0.270$, within $0.6\%$ of the TF value $3/(5\sqrt5)$; Appendix~\ref{app:closure} examines the choice.
The skewness $M_3/S^{3/2}$ of the GP orbits, which measures their departure from a fixed symmetric profile, remains small where the quadratic law of the next subsection applies and grows at larger gain, as Appendix~\ref{app:closure} shows.
With Eqs.~\eqref{eq:Ndot}, \eqref{eq:com}, \eqref{eq:Sdot}, \eqref{eq:Cdot}, and \eqref{eq:closure} we obtain a closed system for five variables,
\begin{equation}
\begin{aligned}
\dot N&=2\gamma NX ,\\
\dot X&=P+2\gamma S ,\qquad \dot P=-X+2\gamma C ,\\
\dot S&=2C ,\qquad \dot C=\frac{q}{S}+\frac{C^2}{S}-S+\frac{gN\kappa}{2\sqrt S} .
\end{aligned}
\label{eq:model}
\end{equation}

The model reproduces the two exact limits.
For $g=0$ and $q=1/4$ it has the solution $S=1/2$, $C=0$, $X=A\sin t$, $N=e^{2\gamma A(1-\cos t)}$, which is Eq.~\eqref{eq:linear}.
For $\gamma=0$ its stationary $S$ is the ground-state value $\sigma_0^2$ by construction, and linearizing Eq.~\eqref{eq:model} about it gives the breathing angular frequency $\omega_B=(3+q/\sigma_0^4)^{1/2}$, the TF value $\sqrt3$ of a one-dimensional cloud~\cite{menotti2002collective} up to the quantum-pressure term.
The model contains no parameter fitted to the orbits; its only inputs are $\sigma_0^2$, a property of the ground state, and $q=1/4$.

\subsection{Shift of the period at small gain}

The mechanism contained in Eq.~\eqref{eq:Xdd} can be followed analytically for small $\gamma$; the steps are given in Appendix~\ref{app:shift}.
To lowest order the COM oscillates harmonically, $X=A\sin t$, and by Eq.~\eqref{eq:Ndot} the atom number varies as $N=e^{2\gamma A(1-\cos t)}\approx1+2\gamma A(1-\cos t)$.
In the static approximation $S$ adjusts to the instantaneous $N$.
A TF cloud has $\mu=(3gN/4\sqrt2)^{2/3}$ and $S=R_{\rm TF}^2/5=2\mu/5$, so that $\delta S/\sigma_0^2=\chi\,\delta N/N$ with $\chi=\partial\ln S/\partial\ln N=2/3$.
Dynamically, however, $S$ behaves as an oscillator of natural angular frequency $\omega_B$ driven at the trap frequency, so the static response is enhanced by the resonance factor $f=\omega_B^2/(\omega_B^2-1)$, equal to $3/2$ in the TF regime; the product $\chi f$ is given in closed form by Eq.~\eqref{eq:chif}.
The oscillating part of $S$ is hence $\delta S=-2\chi f\gamma A\sigma_0^2\cos t$, so that $\dot S=2\chi f\gamma\sigma_0^2X$, and by Eq.~\eqref{eq:Sdot} $C=\dot S/2$.
Inserting into Eq.~\eqref{eq:Xdd} gives $\ddot X=-X+3\gamma\dot S=-(1-6\chi f\gamma^2\sigma_0^2)X$, so that the period grows,
\begin{equation}
\frac{T}{2\pi}=1+3\chi f\,\sigma_0^2\,\gamma^2+O(\gamma^3).
\label{eq:period}
\end{equation}

We write $c_2=(T/2\pi-1)/\gamma^2$, which by Eq.~\eqref{eq:period} approaches $3\chi f\sigma_0^2$ as $\gamma\to0$.
The shift is of second order, and for repulsive interactions it lengthens the period.
At small $\gamma$, $S$ is largest when $N$ is largest, which happens as the cloud leaves the gain side, so that $\dot S$ is in phase with $X$ and the term in $\dot S$ acts against the restoring force.
The resonance factor expresses a level repulsion.
Through the modulation of $N$ the gain couples the dipole mode to the breathing mode, which lies above the dipole mode and pushes the dipole frequency down, as the two-mode relation of Appendix~\ref{app:shift} shows.

The interaction enters Eq.~\eqref{eq:period} through three quantities: the squared width $\sigma_0^2$, which grows with $g$; the static response $\chi$, which vanishes in the noninteracting limit and reaches $2/3$ in the TF regime; and the resonance factor $f$, which moves from $4/3$ to $3/2$ between the same limits.
In the TF regime $\chi f=1$ and $\sigma_0^2=2\mu/5$, which gives the closed form
\begin{equation}
\frac{T}{2\pi}-1\approx\frac65\Big(\frac{3g}{4\sqrt2}\Big)^{2/3}\gamma^2\approx0.79\,g^{2/3}\gamma^2 .
\label{eq:period_TF}
\end{equation}
For $g=100$ it predicts $c_2=16.9$.
Equation~\eqref{eq:period_TF} and the value $\chi f=1$ belong to the one-dimensional equation of state $\mu\propto\rho$.
For $\mu\propto\rho^\zeta$ the breathing frequency is $\omega_B=\sqrt{2+\zeta}$~\cite{menotti2002collective}, and the same argument gives $\chi=2\zeta/(2+\zeta)$, $f=(2+\zeta)/(1+\zeta)$ and $T/2\pi-1=6\zeta\sigma_0^2\gamma^2/(1+\zeta)$.
An elongated three-dimensional condensate in the TF regime, with the gradient along its long axis and lengths in units of the axial oscillator length, has $\zeta=1/2$, so that $T/2\pi-1=2\sigma_0^2\gamma^2=\tfrac47\mu\gamma^2$, with $\sigma_0^2=2\mu/7$ the axial squared width.
A cloud in the TF regime along one transverse direction and frozen in the other has $\zeta=2/3$ and $T/2\pi-1=\tfrac{12}5\sigma_0^2\gamma^2$.

\subsection{The computed orbits}\label{sec:period}

\begin{figure}
\centering
\includegraphics{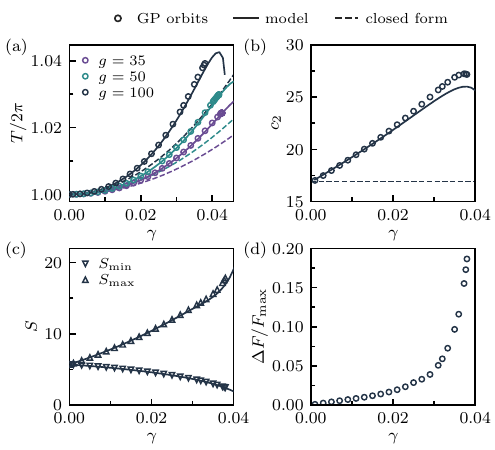}
\caption{The period and the squared width along the orbits for $p_0=10$.
(a) $T$ for $g=35$, $50$ and $100$.
(b)--(d) The $g=100$ family: (b) $c_2$; (c) the smallest and largest $S$ along the orbit; (d) the spread $\Delta F$ over one period of the density profile $F$ rescaled to unit norm and unit squared width, relative to its peak.}
\label{fig:period}
\end{figure}

We compute the orbits numerically, starting from the Kohn orbit at $\gamma=0$ and increasing $\gamma$ in small steps, each computed orbit serving as the initial guess for the next; Appendix~\ref{app:numerics} describes the method.
Figure~\ref{fig:period}(a) shows the period of the symmetric periodic orbits as a function of $\gamma$ for $g=35$, $50$ and $100$ at $p_0=10$.
Each family starts at the Kohn orbit, $T=2\pi$, and its period grows monotonically with $\gamma$, quadratically at small $\gamma$ as Eq.~\eqref{eq:period} requires.

Table~\ref{tab:coef} lists the limit $\gamma\to0$ of $c_2$ for each family, extrapolated from its first computed orbits, at $\gamma\le0.008$, together with three predictions: the closed form of Eq.~\eqref{eq:period_TF}; $3\sigma_0^2$, which is Eq.~\eqref{eq:period} with $\chi f=1$ and $\sigma_0^2$ taken from the numerical ground state instead of the TF value $2\mu/5$; and the five-variable model, for which Eq.~\eqref{eq:period} holds exactly in this limit, with $\chi f$ from Eq.~\eqref{eq:chif}.
Across the factor-of-three range of $g$ considered, the model, the closed form and $3\sigma_0^2$ exceed the GP values by at most $1.2\%$, $2.8\%$ and $5.9\%$, respectively.
Both the closed form and $3\sigma_0^2$ set $\chi f=1$, which overestimates $c_2$ because the quantum pressure lowers $\chi f$ below $1$.
The model includes this reduction, and Appendix~\ref{app:shift} compares its $\chi f$ with the values that the GP orbits imply; the remaining difference comes from the fixed profile of the closure.
In the closed form the error is partly offset by the TF value $2\mu/5$, which lies below $\sigma_0^2$, and all three errors shrink as $g$ grows.

The last three rows change the amplitude instead, at fixed $g=100$, and $c_2$ is the same for $p_0=3$ to $10$ to four digits, confirming that the coefficient of the quadratic law is set by the size of the cloud and not by the amplitude of the oscillation.
As $p_0\to0$ the orbit reduces to the stationary PT-symmetric ground state carrying an infinitesimal dipole excitation; since the limit of $c_2$ does not depend on $p_0$, it equals the coefficient of the $\gamma^2$ shift of the dipole Bogoliubov frequency $\omega_D$ of that state, with $T=2\pi/\omega_D$.
Diagonalizing the Bogoliubov--de Gennes equations about the stationary ground state at $\gamma\le0.008$ reproduces the limits of $c_2$ in Table~\ref{tab:coef} to all digits quoted.

\begin{table}
\caption{\label{tab:coef}The limit $\gamma\to0$ of $c_2$ for the GP orbits, and the predictions of Eq.~\eqref{eq:period}: the TF closed form $6\mu/5$, $3\sigma_0^2$ with $\chi f=1$, and $3\chi f\sigma_0^2$ of the five-variable model.}
\begin{ruledtabular}
\begin{tabular}{ccccc}
$(g,p_0)$ & $6\mu/5$ & $3\sigma_0^2$ & $3\chi f\sigma_0^2$ & $c_2$ \\
\hline
$(35,10)$  & 8.41  & 8.66  & 8.28  & 8.18 \\
$(50,10)$  & 10.67 & 10.88 & 10.57 & 10.47 \\
$(70,10)$  & 13.35 & 13.53 & 13.28 & 13.18 \\
$(100,10)$ & 16.94 & 17.09 & 16.89 & 16.79 \\
$(100,3)$  & 16.94 & 17.09 & 16.89 & 16.79 \\
$(100,5)$  & 16.94 & 17.09 & 16.89 & 16.79 \\
$(100,7)$  & 16.94 & 17.09 & 16.89 & 16.79 \\
\end{tabular}
\end{ruledtabular}
\end{table}

The quadratic law holds only while the modulation of $N$ is weak.
For $g=100$, Fig.~\ref{fig:period}(b) shows the computed $c_2$ rising from the $16.8$ of Table~\ref{tab:coef} to $27.2$ at $\gamma=0.038$, where $N$ varies by a factor $5.5$ over a period and the quadratic law captures only $62\%$ of the shift.
The rise begins linearly in $\gamma$, because $N$ is normalized at $t=0$, where it is smallest, so that the orbit holds on average more atoms than the cloud whose squared width $\sigma_0^2$ enters Eq.~\eqref{eq:period}.
To first order this raises $c_2$ by the relative amount $\tfrac43\gamma A$ in the TF regime, which gives the initial slope of $c_2$ in Fig.~\ref{fig:period}(b), as Appendix~\ref{app:shift} shows.
At larger $\gamma$ the response of $S$ to the strong modulation of $N$ is no longer linear.
The five-variable model, which keeps the full dependence of $S$ on $N$, follows the GP orbits of the $g=100$, $p_0=10$ family; at $\gamma=0.038$ its shift of the period falls short of the GP value by $4\%$, and, as Fig.~\ref{fig:period}(c) shows, it reproduces the range of $S$ along the orbit to within $6\%$.
The loss of accuracy toward the critical gain originates in the closure.
For the rescaled density $F(u)=\sqrt S\,\rho(X+u\sqrt S)/N$ of a GP orbit, the largest variation of $F$ at fixed $u$ over a period, $\Delta F$, shown in Fig.~\ref{fig:period}(d), stays below $2\%$ of the peak value $F_{\max}$ up to $\gamma=0.02$ but reaches $19\%$ at $\gamma=0.038$.
The fixed profile assumed by Eq.~\eqref{eq:closure} thus fails just where the discrepancy grows.

\begin{figure}
\centering
\includegraphics{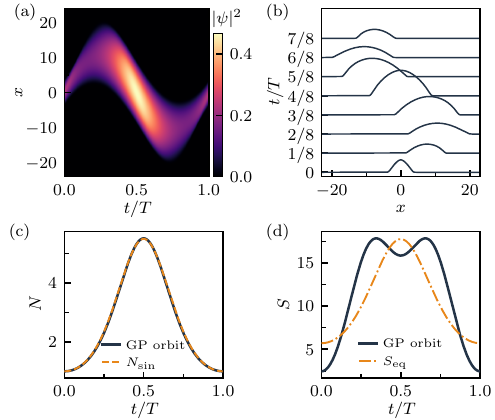}
\caption{The orbit at $\gamma=0.038$ for $g=100$, $p_0=10$.
(a) $|\psi|^2$ over one period and (b) at eight phases, on a common scale.
(c) $N$, with $N_{\rm sin}=\exp[(\gamma AT/\pi)(1-\cos2\pi t/T)]$ for a sinusoidal COM.
(d) $S$, with its equilibrium value $S_{\rm eq}$ at the instantaneous $N$.}
\label{fig:orbit}
\end{figure}

Figure~\ref{fig:orbit} shows an orbit in this regime, at $\gamma=0.038$ for $g=100$.
As Figs.~\ref{fig:orbit}(a)--\ref{fig:orbit}(c) show, the cloud stays single-peaked over the whole period, and $N$ follows closely the modulation $N_{\rm sin}$ that a sinusoidal COM would produce.
The cloud does not follow the equilibrium value of $S$ at its instantaneous atom number, $S_{\rm eq}=\sigma_0^2N(t)^{2/3}$, which Fig.~\ref{fig:orbit}(d) compares with $S$.
The cloud crosses the trap center compressed to $S_0=S(t=0)=0.43\,\sigma_0^2$, the smallest $S$ along the orbit, and $S$ has two maxima per period rather than one, with a local minimum at $t=T/2$, where $N$ is largest.
The split maximum follows from the resonance of Sec.~\ref{sec:ehrenfest}\,C.
The drive lies below $\omega_B$, where $S$ responds in phase, while its second harmonic lies above $\omega_B$, where the response reverses sign; since $S$ varies here by a factor $7$, this linear argument is qualitative.

\section{The critical gain}\label{sec:foldsec}

Each family can be followed only up to a critical gain $\gamma_{\rm c}$.
As $\gamma$ approaches $\gamma_{\rm c}$, the orbits come into resonance with highly excited states of the trap and can become dynamically unstable.
Section~\ref{sec:foldA} describes both effects, and Sec.~\ref{sec:scaling} the dependence of $\gamma_{\rm c}$ on the amplitude of the oscillation and on the size of the cloud.

\subsection{Resonance and stability}\label{sec:foldA}

As $\gamma$ grows, $N$ is modulated more strongly over each period, and the cloud is driven more strongly at the frequency of the orbit and at its harmonics.
The orbits then acquire a small population in highly excited trap levels, through the nonlinear coupling between the condensate and the excited states of the trap that also drives spatiotemporal chaos of a condensate in a harmonic trap~\cite{zhao2025spatiotemporal}.
For $g=100$, $p_0=10$ this happens close to $\gamma=0.0385$, where the fraction of atoms in the trap levels near $n=475$ and $559$ rises by more than two orders of magnitude while remaining below $10^{-3}$, and the density of the cloud itself changes smoothly.
Near this resonance the family turns back toward smaller $\gamma$.
Just below the largest gain that it reaches, two orbits with the same $g$ and $p_0$ therefore exist, and they merge at that gain.
This largest gain is the critical gain $\gamma_{\rm c}$.

The resonant levels are much wider than the cloud.
The trap level $n$ extends to $|x|=\sqrt{2n+1}$, the classical turning point of a particle of the same energy, so that the levels near $n=475$ and $559$ reach $|x|=30.8$ and $33.5$, about three times the amplitude $A$ of the oscillation.
The precise value of $\gamma_{\rm c}$ therefore depends on the trap far from the cloud, and in our computations on the size of the computational domain.
We take as $\gamma_{\rm c}$ of each family the largest gain reached on $|x|\le40$, a domain that contains the resonant levels of the $g=100$, $p_0=10$ family, and quote as its uncertainty the difference from the largest gain reached on the other domains computed, which is at most $4\times10^{-4}$.
For $g=100$, $p_0=10$ this gives $\gamma_{\rm c}=0.03852(6)$.
For the families with $(g,p_0)=(100,7)$, $(50,10)$ and $(35,10)$ a substantial part of the change of the orbit at the turn on $|x|\le40$ lies in states distorted by the edge of the domain rather than in levels of the trap, as Appendix~\ref{app:numerics}\,2 describes.

\begin{figure}
\centering
\includegraphics{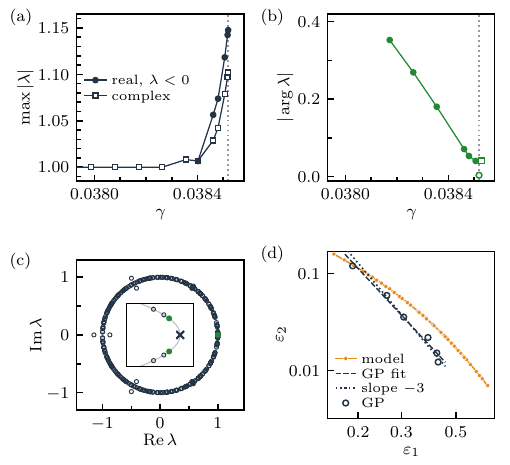}
\caption{The critical gain.
(a)--(c) Floquet multipliers $\lambda$ for $g=100$, $p_0=10$, with $\gamma_{\rm c}$ dotted in (a) and (b).
(a) Largest $|\lambda|$ among the real negative and among the complex multipliers; $|\lambda|>1$ signals an instability.
(b) Phase of the pair that approaches $+1$, filled below the resonance and open within it, at $\gamma_{\rm c}$ (circle) and on a smaller computational domain, on which the family extends slightly beyond $\gamma_{\rm c}$ (square).
(c) The multipliers at $\gamma_{\rm c}$ (open circles), with this pair in green; the inset enlarges $|{\rm Im}\,\lambda|<0.006$ near $+1$, with ${\rm Re}\,\lambda$ stretched by a further factor $400$, and a cross marks the four multipliers that stay at $+1$.
(d) The folds of the model and the critical gains of the six GP families in the plane of Eq.~\eqref{eq:eps}, with a power law fitted to the latter and the slope $-3$ of Eq.~\eqref{eq:foldlaw}.}
\label{fig:fold}
\end{figure}

The linear stability of an orbit is measured by its Floquet multipliers $\lambda$.
A small perturbation of the orbit can be decomposed into Bogoliubov modes of the periodically moving cloud, and after one period each mode returns multiplied by its multiplier.
A mode with $|\lambda|=1$ oscillates, as the Bogoliubov modes of a stable stationary condensate do, whereas $|\lambda|>1$ signals a dynamical instability.
Four multipliers stay at $+1$ on every orbit.
Two of them belong to shifts of the orbit in time and in its global phase, which reproduce the orbit.
The other two belong to changes of $N$ and $p_0$, which carry the cloud onto a neighboring orbit.
That orbit is periodic as well, but its period and phase differ slightly, so that after each period the perturbed cloud falls behind the original one by the same small shift in time and phase, and its distance from the original orbit grows linearly rather than exponentially.
These four multipliers are therefore neutral and do not decide the stability of the orbit, which is set by the remaining ones.
Appendix~\ref{app:numerics}\,3 describes how the multipliers are computed.

For $g=100$, $p_0=10$ the orbits become unstable shortly before $\gamma_{\rm c}$.
As Fig.~\ref{fig:fold}(a) shows, multipliers of the cloud leave the unit circle from $\gamma\approx0.0383$ on, first as a complex quadruplet and then also as a real pair through $-1$.
The real pair belongs to a mode that the orbit drives parametrically and that changes sign from one period to the next, so that the perturbed cloud repeats its shape only every second period; the quadruplet belongs to two modes that the periodic motion couples to each other and amplifies together.
At $\gamma_{\rm c}$ the larger member of the real pair reaches $|\lambda|=1.148$, so that this mode grows by $15\%$ per period, and the largest quadruplet lies $10\%$ off the unit circle.
Other families need not become unstable before $\gamma_{\rm c}$.
At $\gamma_{\rm c}$ of the $p_0=3$ family all multipliers of the cloud lie on the unit circle, whereas for $g=50$, $p_0=10$ a complex quadruplet has left it; Appendix~\ref{app:numerics}\,3 gives these spectra together with that for $g=35$.

Direct time evolution, described in Appendix~\ref{app:numerics}\,3, follows the nonlinear development of the instability.
For $g=100$, $p_0=10$ we add a small perturbation to the orbit at $\gamma_{\rm c}$ and evolve it in time.
The perturbation grows at the rate set by the largest multiplier and saturates after about $100$ periods.
From then on to the $500$th period the evolved state differs by $5$ to $14\%$ from the nearest state on the orbit, and $A$ stays within $1.1\%$ of its value on the orbit.
The instability thus deforms the cloud without destroying the oscillation, and the COM continues to oscillate as on the orbit.
It is also not what makes the family turn back, since the $p_0=3$ family turns back while it is still linearly stable.

The turn itself does not require a multiplier to reach $+1$.
At a given $\gamma$ the orbits are labeled by two parameters, and a family followed at fixed $N(0)$ and $p_0$ can turn back merely because these two parameters cease to determine the orbit uniquely; the two orbits that merge then differ by a step among the orbits at that $\gamma$, which belongs to the neutral multipliers at $+1$.
The five-variable model, whose only internal mode is the breathing mode, turns back in this way, for $g=100$, $p_0=10$ at $\gamma=0.0435$, with the breathing pair at $e^{\pm0.76i}$.
In the GP family one pair of multipliers of the cloud moves along the unit circle toward $+1$ as $\gamma$ approaches $\gamma_{\rm c}$, which means that the frequency of one mode approaches a multiple of the frequency of the orbit.
Its eigenvectors point largely along the family, as Appendix~\ref{app:numerics}\,3 describes, so that near this resonance a small change of $\gamma$ excites this mode strongly.
As Fig.~\ref{fig:fold}(b) shows, the phase of the pair falls from $0.35$ at $\gamma=0.0382$ to $0.04$ at $\gamma=0.03851$, and at $\gamma_{\rm c}$ on $|x|\le40$, whose spectrum Fig.~\ref{fig:fold}(c) shows, the pair lies at $e^{\pm0.003i}$, close to the four multipliers at $+1$.
The resonance with the highly excited states sets in at a slightly smaller gain, and within it the multipliers near $+1$ depend on the computational domain.

\subsection{Position of the critical gain}\label{sec:scaling}

\begin{table}[t]
\caption{\label{tab:fold}Critical gain of six families; $A$ is taken at the critical gain, $S_0^{\min}$ is the smallest $S_0$ along the family, and $\gamma_{\rm fold}^{\rm model}$ is the fold of the five-variable model.
The uncertainty of $\gamma_{\rm c}$ is its spread among the computational domains, as described in Sec.~\ref{sec:foldA}.}
\begin{ruledtabular}
\begin{tabular}{cccccccc}
$g$ & $p_0$ & $R_{\rm TF}$ & $\gamma_{\rm c}$ & $A$ & $S_0^{\min}/\sigma_0^2$ & $\gamma_{\rm c}\sqrt{AR_{\rm TF}}$ & $\gamma_{\rm fold}^{\rm model}$ \\
\hline
100 & 3  & 5.31 & 0.06334(1) & 3.61  & 0.70 & 0.277 & 0.0640 \\
100 & 5  & 5.31 & 0.0524(4) & 5.69  & 0.59 & 0.288 & 0.0561 \\
100 & 7  & 5.31 & 0.0438(2) & 7.63  & 0.56 & 0.279 & 0.0502 \\
100 & 10 & 5.31 & 0.03852(6) & 10.86 & 0.41 & 0.293 & 0.0435 \\
50  & 10 & 4.22 & 0.04185(2) & 10.63 & 0.38 & 0.280 & 0.0493 \\
35  & 10 & 3.74 & 0.0424(4) & 10.48 & 0.40 & 0.265 & 0.0524 \\
\end{tabular}
\end{ruledtabular}
\end{table}

The critical gain depends on the amplitude of the oscillation and on the size of the cloud.
Table~\ref{tab:fold} lists $\gamma_{\rm c}$ for six families: four at $g=100$ with $p_0$ from $3$ to $10$, and two more at $p_0=10$, with $g=50$ and $35$, together with $A$ at the critical gain and the smallest $S_0$ reached along the family.
The critical gain rises as $p_0$ is reduced, and depends much more weakly on $g$.
A fit with free exponents, $\gamma_{\rm c}\propto A^{-\nu_A}R_{\rm TF}^{-\nu_R}$, gives $\nu_A=0.46(3)$ and $\nu_R=0.27(8)$, so that the dependence on $A$ is close to the inverse square root, whereas $R_{\rm TF}$, which varies only by a factor $1.4$ among the families, enters more weakly.
The six values of $\gamma_{\rm c}\sqrt{AR_{\rm TF}}$ nevertheless lie within $6\%$ of their mean,
\begin{equation}
\gamma_{\rm c}\sqrt{A\,R_{\rm TF}}=0.280 ,
\label{eq:foldlaw}
\end{equation}
with a standard deviation of $0.009$, so that Eq.~\eqref{eq:foldlaw} describes the computed critical gains to $3\%$, although it gives $R_{\rm TF}$ a larger exponent than the free fit.
It is an empirical relation for $A/R_{\rm TF}$ from $0.7$ to $2.8$ and cannot be continued to small amplitude.
As $A\to0$ the orbit reduces to the stationary PT-symmetric ground state, which for $g=100$ exists only up to $\gamma=0.163$, whereas Eq.~\eqref{eq:foldlaw} diverges; at that threshold its chemical potential is $\mu_c=16.98$, so that $\gamma\sqrt{2\mu_c}=0.95$, close to the value $0.93$ of this combination in the TF limit~\cite{gallo2014thomas}.

The mechanism of Sec.~\ref{sec:ehrenfest} involves two dimensionless combinations of the gain, the amplitude and the size of the cloud,
\begin{equation}
\varepsilon_1=\gamma p_0 ,\qquad \varepsilon_2=\frac{\gamma\sigma_0^2}{p_0} .
\label{eq:eps}
\end{equation}
The first sets the depth of the atom-number modulation, $N_{\max}/N_{\min}\simeq e^{4\gamma A}$ with $A\simeq p_0$, and the second the strength with which a given modulation displaces the COM through the term $2\gamma S$ in Eq.~\eqref{eq:com}.
Rescaling Eq.~\eqref{eq:model} by $X=p_0\bar X$, $P=p_0\bar P$, $S=\sigma_0^2\bar S$, $C=\sigma_0^2\bar C$ and eliminating $g$ with the ground-state condition of Appendix~\ref{app:closure} shows that $\varepsilon_1$ and $\varepsilon_2$ are the only parameters of the model, apart from the quantum-pressure term $q/\sigma_0^4$, which does not exceed a few percent over the range of $g$ considered.
The folds of the model must therefore lie close to a single curve in the $(\varepsilon_1,\varepsilon_2)$ plane, and Fig.~\ref{fig:fold}(d) shows that fifty folds computed for $g$ from $25$ to $280$ and $p_0$ from $3$ to $20$ indeed do.

For the GP equation the same reduction follows from dimensional analysis of the hydrodynamic equations once the quantum pressure is dropped, since scaling lengths by $R_{\rm TF}$ leaves $\gamma R_{\rm TF}$ and $p_0/R_{\rm TF}$ as the only parameters, and these fix $\varepsilon_1$ and $\varepsilon_2$.
This reduction covers the dynamics of the cloud, but not the highly excited trap levels whose resonance ends the GP families, since at these energies the excitations behave as free particles, whose dispersion arises from the quantum pressure that the reduction drops.
In Fig.~\ref{fig:fold}(d) the six GP families trace a curve steeper than that of the model; a power law fitted to them has slope $-2.7(2)$, close to the $-3$ that Eq.~\eqref{eq:foldlaw} gives with $A\simeq p_0$ and $R_{\rm TF}^2\simeq5\sigma_0^2$.
A test of the reduction for $\gamma_{\rm c}$ would require families with the same $p_0/R_{\rm TF}$ and different $g$, whereas each of the six families has its own $p_0/R_{\rm TF}$; Appendix~\ref{app:fits} discusses the fits.

The five-variable model describes all six families comparably well.
Up to $0.9\,\gamma_{\rm c}$ its shift of the period falls short of that of the GP orbits by at most $11\%$, the largest deficit occurring at $p_0=3$, and the smallest and largest $S$ along the orbit agree with those of the GP orbits to within $7\%$.
Closer to $\gamma_{\rm c}$ the two descriptions part.
The model carries the compression of the cloud much further than the GP orbits do; at the fold of its $g=100$, $p_0=10$ family $S_0=0.070\,\sigma_0^2$, whereas the GP family reaches its critical gain while $S_0$ is still above $0.41\,\sigma_0^2$.
The closure of Eq.~\eqref{eq:closure} keeps the profile fixed however strongly the cloud is compressed, whereas the GP cloud develops internal structure as it narrows.
As Table~\ref{tab:fold} shows, the fold of the model lies above $\gamma_{\rm c}$ for every family, by $1\%$ at $p_0=3$ and by $7$ to $24\%$ for the other five.

\section{Experimental realization}\label{sec:experiment}

Two platforms could provide a harmonic trap with a linear gain-loss profile.

In the paraxial approximation a monochromatic beam of vacuum wave number $k_0$ propagating along $z$ in a medium of index $n_0+\delta n(x)$ obeys $i\partial_zE=-(2k)^{-1}\partial_x^2E-k_0\,\delta n\,E$, with $k=n_0k_0$ and $z$ in the role of time.
A transverse parabolic profile $\delta n=-\tfrac12n_0\Omega^2x^2$ is a graded-index waveguide in which a beam oscillates with period $z_T=2\pi/\Omega$, an imaginary part $\delta n=-in_0\beta x$ is a gain-loss gradient with gain where $\beta x>0$, and a defocusing nonlinearity supplies the term $g|E|^2$.
With the beam confined along the second transverse direction strongly enough that the nonlinearity does not change its profile there, scaling $x$ by $(k\Omega)^{-1/2}$ and $z$ by $1/\Omega$ maps the problem onto Eq.~\eqref{eq:gpe} with $\gamma=\beta k^{1/2}\Omega^{-3/2}$.
The Kerr response of a warm vapor~\cite{fontaine2018observation} saturates at high intensity and is nonlocal over the distance the atoms travel during their response time, so that the cubic term applies to intensities well below saturation and to structures of the beam, down to the healing length, larger than that distance.

Periodic gain-loss profiles have been imprinted optically in a warm rubidium vapor~\cite{zhang2016observation}.
There, standing-wave coupling and pump fields form two lattices whose relative displacement tunes the real and imaginary parts of the susceptibility, the gain being supplied by active Raman amplification, and the onset of PT symmetry breaking has been observed in the resulting array of gain and loss waveguides.
The present proposal replaces that periodic profile by a parabolic real part and a linear imaginary part.

At the critical gain the intensity gain and loss coefficients at the turning points reach $2\gamma_{\rm c}A\,\Omega\approx0.8\,\Omega$, one to a few inverse centimeters for $z_T$ between $2$ and $5$~cm.
For $g=100$ and a wavelength of $780$~nm the nonlinear index change at the center of the ground state, $\mu\Omega/k_0$, is then $2.2$ to $5.5\times10^{-4}$, and along the orbit at $\gamma_{\rm c}$ it reaches about $3.5$ times this value; the real index contrast at the turning points, $\tfrac12A^2\Omega/k_0$, is $0.9$ to $2.3\times10^{-3}$; a longer $z_T$ lowers both, and a smaller $g$ the first.
Background absorption of the vapor acts as a uniform loss, which lowers the nonlinearity along the cell and has to be compensated by the gain.
The observables are the transverse density and the total power at the exit face, the latter being the direct counterpart of Eq.~\eqref{eq:Ndot}; since only that face is accessible, the dependence on $z$ is reconstructed by using cells of different lengths.

For atoms the loss side can be realized directly.
A resonant beam or an electron beam whose intensity grows linearly with the distance from the center removes atoms at the required rate; spatially resolved loss is an established tool~\cite{labouvie2016bistability}, and engineered loss has been used to give a condensate an imaginary gauge potential~\cite{tao2026imaginary}.
With $^{87}$Rb in a trap of a few tens of hertz the critical gain corresponds to a loss rate of order $10^2$~s$^{-1}$ at the turning point, a few tens of micrometers from the center.
The one-dimensional equation applies if the chemical potential lies below $\hbar\omega_\perp$, with $\omega_\perp$ the transverse trap frequency; for $g=100$ it is $14\hbar\omega$ in the ground state, and along the orbit at $\gamma_{\rm c}$ the interaction energy at the center of the cloud reaches about $50\hbar\omega$, so that $\omega_\perp\ge100\,\omega$ is required.
The shift of the period is set within the cloud, whereas for $g=100$, $p_0=10$ the critical gain is set by trap levels near $n=500$, about $500\hbar\omega$ above the ground state and extending to $|x|\approx30$, far beyond the cloud; in an experiment $\gamma_{\rm c}$ therefore also depends on the anharmonicity of the trap far from the cloud and on transverse excitations at these energies.

Loss alone does not suffice, however.
A lossy profile $i\gamma(x-x_c)$, with $x_c$ beyond the edge of the cloud at the turning points so that no gain is needed, differs from Eq.~\eqref{eq:gpe} by a uniform decay that multiplies $gN$ by $e^{-2\gamma x_cT}$ per period, about $4\times10^{-5}$ for $x_c=20$, $\gamma=\gamma_{\rm c}$ and $T=2.08\pi$.
The passive schemes used to emulate PT symmetry with atoms~\cite{li2019observation} would therefore remove the interaction that causes the effect, unless the scattering length were raised by the same factor.
A displacement $\delta$ of the center of the gain profile from the trap center acts in the same way, multiplying $N$ by $e^{-2\gamma\delta T}$ per period; since the orbits form a continuous family in $N$, this loss accumulates, and over $M$ periods $\delta$ must remain much smaller than $(2\gamma TM)^{-1}$, about $0.02$ for $M=100$, or $50$~nm for $^{87}$Rb at $\omega=2\pi\times20$~Hz.
The gain side requires a reservoir that feeds the condensate coherently at a rate growing linearly with $x$; reservoir wells coupled to a double well have been proposed to supply balanced gain and loss~\cite{kreibich2014realizing}, and continuous loading of a condensate from a reservoir by Bose-stimulated gain has been demonstrated~\cite{chen2022continuous}.

\begin{figure}
\centering
\includegraphics{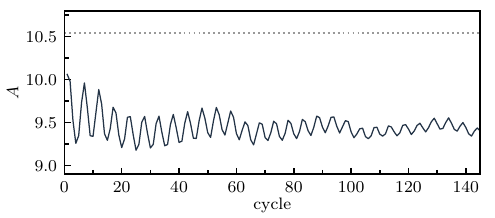}
\caption{Amplitude of successive cycles of the ground state kicked to $\dot X(0)=10$ at $\gamma=0.035$ for $g=100$, a cycle running from one crossing of the center toward the gain side to the next; the dotted line marks $A$ of the orbit at the same gain.}
\label{fig:kick}
\end{figure}

The oscillation can be started by displacing or kicking the ground state, which does not lie on the orbit, since $S_0<\sigma_0^2$ along the family; in the fluid of light, whose cell of $10$~cm holds only two to five periods, the input beam can instead be shaped to the state of the orbit at $t=0$.
At $\gamma=0.035$ and $g=100$ we follow over $150$ trap periods the ground state of the trap at $\gamma=0$, kicked by $e^{iP(0)x}$ to $\dot X(0)=p_0=10$, with $P(0)=\dot X(0)-2\gamma\sigma_0^2$ by Eq.~\eqref{eq:com}.
As Fig.~\ref{fig:kick} shows, after a transient of a few tens of cycles the amplitude fluctuates about $9.4$, below $A=10.5$ of the orbit, while $N$ at the crossings stays near $1.07$.
From the $30$th cycle on, the period of individual cycles scatters by $0.2\%$ rms about a mean that equals $T$ of the orbit to within $2\%$ of the shift $T-2\pi$.
For $\gamma=0.01$ the relative shift of the period is about $2\times10^{-3}$, so that the dipole period has to be resolved to better than $10^{-3}$.

\section{Conclusion}\label{sec:conclusion}

A PT-symmetric gain-loss gradient alone leaves the period of the Kohn oscillation of a trapped condensate intact~\cite{longhi2016pt}.
Interactions break this protection because the gain and the loss change $N$ and with it the interaction energy, through a chain of couplings in which the modulated $N$ drives $S$, and $S$ acts back on the COM through $\dot S$ and the covariance $C$.
At small gain the relative shift of the period is $0.79\,g^{2/3}\gamma^2$ in the TF regime, with no adjustable parameter; at vanishing amplitude it corresponds to the lowering of the dipole Bogoliubov frequency of the stationary ground state.
A five-variable moment model follows the period and the width of the cloud along the orbits nearly up to the critical gain.

The family can be followed up to a critical gain, where its orbits come into resonance with highly excited states of the trap and the family turns back; over the six computed families the critical gain decreases with the amplitude and the size of the cloud, and $\gamma_{\rm c}\sqrt{AR_{\rm TF}}$ stays within $6\%$ of $0.280$.
For the dynamics of the cloud in the TF regime, dimensional analysis of the hydrodynamic equations shows that the gain, the amplitude and the size of the cloud enter only through $\varepsilon_1$ and $\varepsilon_2$, and the folds of the model obey this reduction closely.
Because the resonant states reach far beyond the cloud, the precise value of the critical gain depends on the trap there, and in our computations on the computational domain, by up to $4\times10^{-4}$, about $1\%$.
For $g=100$, $p_0=10$ the orbits become dynamically unstable shortly before the critical gain, but the instability does not destroy the oscillation, and the COM continues to oscillate as on the orbit.

The family that starts from the first excited state of the trap, the analog of a dark soliton, can be treated with the same method; stationary nonlinear modes that branch from the excited states of this potential are known~\cite{zezyulin2012nonlinear}, and a gain-loss profile turns a dark soliton gray~\cite{achilleos2012solitons}.
In two dimensions the gradient selects one axis, and the interplay with vortices and with the transverse breathing mode is open.
The mean-field description assumes throughout that the atoms added on the gain side join the condensate coherently, which the fluid of light realizes directly and which a continuously loaded condensate~\cite{chen2022continuous} would put to the test.

\section*{Data availability}
The data that support the findings of this article are openly available~\cite{meng2026data}.
The scripts that produce the figures and the tables are available from the corresponding author upon reasonable request.

\begin{acknowledgments}
This work was partially supported by the Hebei Natural Science Foundation (A2026502021), the Fundamental Research Funds for the Central Universities (2026MS219), and the Research and Practice Project of Higher Education Teaching Reform of Hebei Province (Nos. 2025GJJG412 and 2026GJJG415).
\end{acknowledgments}

\appendix
\numberwithin{equation}{section}

\section{Moment hierarchy}\label{app:moments}

We use $\xi=x-X$ and $\eta=p-P$ for the centered operators and $D=\tfrac12(xp+px)$, so that $C=\langle D\rangle-XP$.

For $O=x$ the commutator is $[H_R,x]=-ip$ and $\{W,x\}=2\gamma x^2$, so Eq.~\eqref{eq:ehrenfest} gives $\dot X=P+2\gamma(\langle x^2\rangle-X^2)=P+2\gamma S$.
For $O=p$ we have $i[H_R,p]=-\partial_x(\tfrac12x^2+g\rho)$ and $\{W,p\}=\gamma(xp+px)$, hence
\begin{equation}
\dot P=-X-g\langle\partial_x\rho\rangle+2\gamma(C+XP)-2\gamma XP=-X+2\gamma C ,
\end{equation}
because $\langle\partial_x\rho\rangle=\tfrac1{2N}\int\partial_x(\rho^2)dx=0$ for a density that vanishes at infinity.
This expresses the absence of a net interaction force on the COM, and it holds for any local nonlinearity, not only the cubic one.

For $O=x^2$, $i[H_R,x^2]=xp+px$ and $\{W,x^2\}=2\gamma x^3$, so
\begin{equation}
\frac{d\langle x^2\rangle}{dt}=2(C+XP)+2\gamma\big(\langle x^3\rangle-X\langle x^2\rangle\big).
\end{equation}
Subtracting $2X\dot X$ and using $\dot X=P+2\gamma S$,
\begin{equation}
\dot S=2C+2\gamma\big(\langle x^3\rangle-3X\langle x^2\rangle+2X^3\big)=2C+2\gamma M_3 ,
\end{equation}
with $M_3=\langle\xi^3\rangle$, so that the gain enters the equation for $S$ only through the third central moment.

For $O=D$ the commutator is $i[H_R,D]=p^2-x^2-gx\,\partial_x\rho$, whose expectation value is $\langle p^2\rangle-\langle x^2\rangle-g\langle x\partial_x\rho\rangle$, and, using $\int x\,\partial_x(\rho^2)dx=-\int\rho^2dx$ after an integration by parts, $-g\langle x\partial_x\rho\rangle=\tfrac12gN\!\int\tilde\rho^2dx$.
The anticommutator simplifies to $\{W,D\}=2\gamma\,xpx$, since $\tfrac12(x^2p+px^2)=xpx$.
Collecting the terms and subtracting $\dot XP+X\dot P$ gives Eq.~\eqref{eq:Cdot} with
\begin{equation}
M_{21}=\langle xpx\rangle-2XC-PS-PX^2=\langle\xi\eta\xi\rangle ,
\end{equation}
the symmetrized third central moment of position and momentum.

\section{Closure and its constants}\label{app:closure}

The closure assumes a profile that keeps its shape while its width changes,
\begin{equation}
\tilde\rho(x,t)=S^{-1/2}F\!\big(\xi/\sqrt S\big),\qquad \int\!F\,du=\int\!u^2F\,du=1 ,
\end{equation}
and a phase quadratic in $\xi$, $\varphi=Px+(C/2S)\xi^2$, whose gradient $\partial_x\varphi=P+(C/S)\xi$ is the velocity field that produces the covariance, $\tfrac12\langle\xi\eta+\eta\xi\rangle=(C/S)\langle\xi^2\rangle=C$.
Writing $\psi=\sqrt\rho\,e^{i\varphi}$ separates $\langle p^2\rangle$ into an amplitude and a phase contribution,
\begin{equation}
\langle p^2\rangle=\int\Big[\big(\partial_x\sqrt{\tilde\rho}\big)^2+\tilde\rho\,(\partial_x\varphi)^2\Big]dx ,
\end{equation}
of which the first gives $q/S$ under the scaling of $F$ and the second $P^2+C^2/S$, the cross term vanishing because $\langle\xi\rangle=0$.
The two relations of Eq.~\eqref{eq:closure} follow with
\begin{equation}
\kappa=\int\!F^2du ,\qquad q=\int\!\big[(\sqrt F)'\big]^2du ,
\end{equation}
the second being the quantum pressure of the profile.
For the inverted parabola of the TF limit, $F(u)=\tfrac{3}{4\sqrt5}(1-u^2/5)$ on $|u|<\sqrt5$, $\kappa=3/(5\sqrt5)=0.2683$.
We instead fix $\kappa$ by requiring the correct ground-state squared width, $q/\sigma_0^2-\sigma_0^2+g\kappa/(2\sigma_0)=0$ with $\sigma_0^2$ taken from the numerical ground state, which gives $\kappa=0.2698$ for $g=100$.
The quantum pressure of the strict parabola diverges at its edge, so $q$ cannot be taken from $F$ in the TF limit, and we keep the Gaussian value $q=1/4$.
Along the orbits the term $q/S$ remains small compared with the interaction term $gN\kappa/2\sqrt S$.
At the smallest $S$ that the GP orbits of the $g=100$, $p_0=10$ family reach, $0.41\,\sigma_0^2$, it amounts to $1\%$ of that term, and at the fold of the model, where $S_0=0.070\,\sigma_0^2$, to $3\%$.

A fully Gaussian variational closure, with $\kappa=1/(2\sqrt\pi)$ and $q=1/4$ taken from the same profile and the ground-state squared width determined by the model, moves the folds for $(g,p_0)=(100,3)$, $(100,10)$ and $(35,10)$ by less than $2\%$, so that the discrepancy with the GP orbits originates in the assumption of a fixed profile, which fails as the cloud is compressed and deformed, rather than in the mixed origin of the two constants.
The skewness $M_3/S^{3/2}$ of the GP orbits grows accordingly along the family; for $g=100$, $p_0=10$ its largest value along the orbit is $0.04$ at $\gamma=0.029$ and $0.19$ at $\gamma=0.038$.

\section{Shift of the period at small gain}\label{app:shift}

Take $X=A\sin t$ at lowest order, so that Eq.~\eqref{eq:Ndot} gives $N=e^{2\gamma A(1-\cos t)}\simeq1+2\gamma A(1-\cos t)$.
Linearizing the equations for $S$ and $C$ in Eq.~\eqref{eq:model} about $(S,C,N)=(\sigma_0^2,0,1)$, with the coefficient of $\delta N$ rewritten by the ground-state condition of Appendix~\ref{app:closure}, gives
\begin{equation}
\delta\ddot S=-\omega_B^2\,\delta S+2\Big(\sigma_0^2-\frac{q}{\sigma_0^2}\Big)\delta N ,\qquad \omega_B^2=3+\frac{q}{\sigma_0^4} .
\end{equation}
The oscillating part of the drive has the trap frequency, which is unity in our units, and the response to it is
\begin{equation}
\delta S=\chi f\,\sigma_0^2\,\delta N,\qquad
\chi f=\frac{2\big(1-q/\sigma_0^4\big)}{2+q/\sigma_0^4} ,
\label{eq:chif}
\end{equation}
the static response $\chi=\partial\ln S/\partial\ln N$ multiplied by the resonance factor $f=\omega_B^2/(\omega_B^2-1)$ of a driven oscillator.
In the TF limit $\chi=2/3$ and $f=3/2$, so $\chi f=1$ exactly.
At the trap frequency a cloud in the TF regime thus responds to a modulation of $N$ as if $S$ were simply proportional to $N$, the resonant enhancement making up exactly for the static response being $2/3$.
For $g=35$, $50$, $70$ and $100$, Eq.~\eqref{eq:chif} gives $\chi f=0.956$, $0.972$, $0.982$ and $0.988$, whereas the GP values of $c_2$ in Table~\ref{tab:coef} correspond to $\chi f=c_2/3\sigma_0^2=0.944$, $0.962$, $0.974$ and $0.983$; the difference comes from the fixed profile of the closure.

Inserted into Eq.~\eqref{eq:Xdd} as in Sec.~\ref{sec:ehrenfest}\,C, the response of Eq.~\eqref{eq:chif} gives Eq.~\eqref{eq:period}.
One factor of $\gamma$ comes from the modulation of $N$ in Eq.~\eqref{eq:Ndot} and the other from the coupling of $S$ back to the COM in Eq.~\eqref{eq:com}, while the response of $S$ to $N$ between them vanishes with the interaction, so that the shift is absent both for $\gamma=0$ and for $g=0$.

The same result follows from the coupled linear equations of the two modes.
With $\delta\dot N=2\gamma X$, $\delta\ddot S+\omega_B^2\delta S=\omega_B^2\chi\sigma_0^2\delta N$ and $\ddot X=-X+3\gamma\,\delta\dot S$, a solution $\propto e^{i\nu t}$ requires
\begin{equation}
(1-\nu^2)(\omega_B^2-\nu^2)=6\chi\sigma_0^2\omega_B^2\gamma^2 ,
\end{equation}
whose root near $\nu=1$ reproduces Eq.~\eqref{eq:period} to order $\gamma^2$.
The coupling lowers the dipole frequency and raises the breathing frequency, which is the level repulsion of Sec.~\ref{sec:ehrenfest}\,C.

The constant part of $\delta N$ shifts the mean of $S$ at $O(\gamma)$ and does not enter $\dot S$ at this order.
It does so at the next order.
Averaged over a period the orbit holds $1+2\gamma A$ times as many atoms as at $t=0$, where $N$ is normalized, and to this order its $c_2$ is that of a cloud with this larger number of atoms, which exceeds the limit of Eq.~\eqref{eq:period} by the relative amount $2\gamma A\,\partial\ln(\chi f\sigma_0^2)/\partial\ln g$, or $\tfrac43\gamma A$ in the TF regime; the remainder in Eq.~\eqref{eq:period} is therefore of third order in $\gamma$.
For $g=100$, $p_0=10$ this correction gives $dc_2/d\gamma\approx2.3\times10^2$ as $\gamma\to0$, the initial slope of $c_2$ in Fig.~\ref{fig:period}(b).

\section{Numerical methods}\label{app:numerics}

\subsection{Periodic orbits from half-period shooting}

A periodic orbit of Eq.~\eqref{eq:gpe} is a state $\psi_0$ and a period $T$ such that $\mathcal U_T\psi_0=e^{i\theta}\psi_0$, where $\mathcal U_t$ denotes the nonlinear evolution over a time $t$.
We look for orbits that are mapped onto themselves by the reversing operation $\mathcal R\psi(x)=\psi^*(-x)$ of Sec.~\ref{sec:model}.
Such an orbit passes through the set $\mathrm{Fix}(\mathcal R)$ of PT-symmetric states at $t=0$, which we take as the crossing of the trap center, and again at $t=T/2$, up to the global phase~\cite{lamb1998time}.
The second passage can be written as $e^{-i\alpha}\mathcal U_{T/2}\psi_0\in\mathrm{Fix}(\mathcal R)$ for some phase $\alpha$.
Since $\mathcal R\,\mathcal U_t\,\mathcal R=\mathcal U_{-t}$, these two conditions imply the full-period condition with $\theta=2\alpha$~\cite{lamb1998time}.
The shooting problem therefore involves only half a period, which halves both the cost of an iteration and the number of unknowns.

On a grid of an odd number $N_x$ of points symmetric about $x=0$, a state in $\mathrm{Fix}(\mathcal R)$ has a real even part and an imaginary odd part and is described by $N_x$ real numbers.
The unknowns are these $N_x$ numbers together with $T$ and $\alpha$.
The equations are the $N_x$ real components of $e^{-i\alpha}\mathcal U_{T/2}\psi_0$ outside $\mathrm{Fix}(\mathcal R)$, which must vanish, together with the two conditions that fix the family, $N(0)=1$ and $\dot X(0)=P(0)+2\gamma S_0=p_0$, the latter being Eq.~\eqref{eq:com} evaluated at the crossing.
The count of unknowns and equations is thus balanced, the time-translation and phase symmetries of the orbit are removed by the restriction to $\mathrm{Fix}(\mathcal R)$ rather than by additional constraints, and the Jacobian of the system is regular along the family except where the family turns back or branches.

The Jacobian is computed exactly by tangent propagation, in which the $N_x$ unit perturbations of $\psi_0$ are propagated along the reference trajectory by the linearized evolution, while the columns of $T$ and $\alpha$ follow from the same trajectory.
Its smallest singular values, which belong to short-wavelength Bogoliubov modes that the half-period map returns nearly to $\mathrm{Fix}(\mathcal R)$, lie three orders of magnitude below the typical ones, and we therefore use a Levenberg--Marquardt step constructed from the singular value decomposition of the Jacobian.

\subsection{Continuation}

The family of orbits is followed in $\gamma$ starting from the Kohn orbit at $\gamma=0$, which at $t=0$ is the ground state multiplied by $e^{ip_0x}$.
Away from the critical gain we use natural continuation, in which the previous orbits, extrapolated linearly, serve as the initial guess at the next value of $\gamma$.
If the iteration fails at a given $\gamma$ the step is first doubled, so as to pass over the narrow windows in which a high-order Bogoliubov resonance contracts the basin of attraction, and is halved only if two enlarged steps also fail.
The orbits so obtained return to themselves after one full period to within $2\times10^{-5}$ in relative norm, nine in ten of them to within a few times $10^{-7}$.
Close to the critical gain we switch to pseudo-arclength continuation~\cite{seydel2010practical}, in which $\gamma$ becomes an unknown and each new orbit is required to lie in the hyperplane orthogonal to the tangent of the family through the predicted point.
This form of continuation is insensitive to the vertical tangent where the family turns back; it passes the turn, which is located by the change of sign of $d\gamma/ds$, with $s$ the arclength along the family.
On the orbits that it computes the Euclidean norm of the residual is below $10^{-6}$.

Near the critical gain the orbits come into resonance with highly excited states, and which states these are depends on the computational domain.
We identify them by expanding the orbit in the eigenstates of the discretized Hamiltonian $-\tfrac12\partial_x^2+\tfrac12x^2$, whose eigenvalues agree with $n+\tfrac12$ to within $10^{-3}$ for $n<492$, $776$ and $1126$ on $|x|\le32$, $40$ and $48$ at $\Delta x=0.0625$; the higher eigenstates are distorted by the edge of the domain or by the cutoff of the grid and are not levels of the trap.
For $g=100$, $p_0=10$ the continuation turns at $\gamma=0.038519$ on $|x|\le40$ with $N_x=1281$, where the orbit mixes with the trap levels near $n=475$ and $n=559$.
It turns at $\gamma=0.038529$ on $|x|\le32$ with $N_x=1025$, where distorted states take the place of the levels near $n=559$, and at $\gamma=0.038535$ on $|x|\le48$ with $N_x=1537$, where the orbit also takes up states near the cutoff of the grid.

On a Fourier grid, however, the gain profile $\gamma x$ jumps from $\gamma L$ to $-\gamma L$ across the periodic boundary of the domain $|x|\le L$, and this produces spurious modes that grow in time, as discussed in Appendix~\ref{app:numerics}\,3.
We therefore repeat the continuation with the gain tapered smoothly to zero near the edge, $\gamma x$ being replaced by $\gamma x\,h(x)$ with $h=1$ for $|x|\le x_a$, $h=0$ for $|x|\ge x_b$, and $h=1-10r^3+15r^4-6r^5$ in between, $r=(|x|-x_a)/(x_b-x_a)$, with $(x_a,x_b)=(22,30)$ on $|x|\le32$ and $(24,34)$ on $|x|\le40$.
Over a period the orbits at the turn on $|x|\le32$ and $40$ have less than $10^{-4}$ of their atoms beyond $|x|=22$, the taper preserves PT symmetry and reversibility, and it returns every multiplier of the linear problem to the unit circle.

With the tapered gain the continuation on $|x|\le40$ turns at $\gamma=0.038516$, $3\times10^{-6}$ below the turn without the taper.
On $|x|\le32$ the tapered continuation agrees with that on $|x|\le40$ to $2\times10^{-6}$ in $T/2\pi$ up to $\gamma=0.03851$; beyond, it takes up states of the domain and does not turn, and at $\gamma=0.03858$, where its corrector ceases to converge, these states hold $2\times10^{-3}$ of the atoms.

\begin{figure}
\centering
\includegraphics{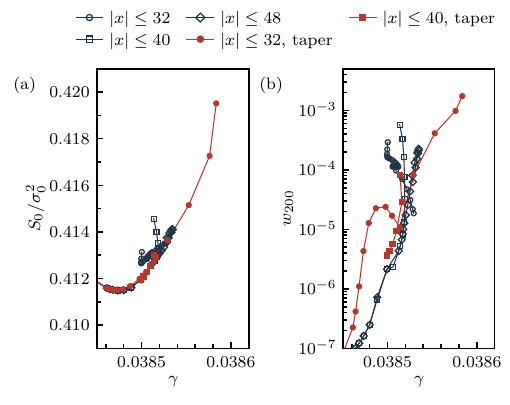}
\caption{The continuation of the family for $g=100$, $p_0=10$ in five computations.
(a) $S_0$ evaluated within $|x|<12$.
(b) The weight $w_{200}$ of the orbit in the trap levels $n\geq200$.}
\label{fig:end}
\end{figure}

Figure~\ref{fig:end} compares these five computations of the $g=100$, $p_0=10$ family.
In each of them the weight $w_{200}$ of the orbit in the trap levels $n\ge200$ rises steeply close to $\gamma=0.0385$, whereas the density of the cloud changes smoothly, and in the four computations that turn it stays below $6\times10^{-4}$ past the turn.

For all six families we take the gain of the turn on $|x|\le40$ as $\gamma_{\rm c}$ and quote as its uncertainty the difference from the largest gain reached on the other domains, with a minimum of one unit in the last digit.
For the five families other than $(100,10)$ the other domain is $|x|\le32$, on which the natural continuation reaches $\gamma=0.06334$, $0.05273$, $0.04399$, $0.04187$ and $0.04278$ for $(g,p_0)=(100,3)$, $(100,5)$, $(100,7)$, $(50,10)$ and $(35,10)$; for $(100,5)$, $(100,7)$ and $(35,10)$ these orbits lie well beyond the turn on $|x|\le40$ and close to within $10^{-8}$.
For $(100,7)$, $(50,10)$ and $(35,10)$, $24$ to $69\%$ of the tangent of the family at the turn on $|x|\le40$ lies in the states $n\ge776$, which are states of the domain rather than levels of the trap.

\subsection{Floquet multipliers}

The linear stability of an orbit is determined by the monodromy matrix of the full-period map $e^{-2i\alpha}\mathcal U_T$, linearized at $\psi_0$ and acting on the real vector $(\mathrm{Re}\,\delta\psi,\mathrm{Im}\,\delta\psi)$, since the nonlinearity is not complex linear.
The $2N_x$ columns are obtained by tangent propagation over one period, and all multipliers are computed by direct diagonalization.
For a symmetric orbit reversibility implies that multipliers come in pairs $(\lambda,1/\lambda)$, and reality adds the complex conjugates, so multipliers lie on the unit circle or leave it in real pairs or complex quadruplets~\cite{lamb1998time}; a symmetric orbit is therefore either neutrally stable or unstable, and never attracting.

Four multipliers sit at $+1$.
Two eigenvectors are generated by the continuous symmetries, namely the direction along the orbit and the direction of the global phase, while for $\gamma\neq0$ the two parameters of the orbits supply generalized eigenvectors, since displacing $p_0$ or $N$ alters the period and the accumulated phase $\alpha$, so that after one period the two displacements acquire components along the orbit and along the phase direction.
The spectrum at $+1$ therefore has algebraic multiplicity four and geometric multiplicity two, in two Jordan blocks of size two.
In the computed spectra these four multipliers lie within $4\times10^{-4}$ of $+1$ at $\gamma=0.029$ and within $10^{-4}$ at the turn on $|x|\le40$.

Any finite computational domain supports high-lying modes that are absent from the problem on the infinite line.
In the linear problem the jump of the gain profile across the periodic boundary of the Fourier grid takes some of them off the unit circle, whereas the trap on the line has a real spectrum for every $\gamma$.
At $\gamma=0.0385$ the largest multiplier of the linear map over the period $T=2.08\pi$ has modulus $1.015$ on $|x|\le32$, $1.024$ on $|x|\le40$ and $1.47$ on $|x|\le48$, and all multipliers return to the unit circle with the tapered gains of Appendix~\ref{app:numerics}\,2 on $|x|\le32$ and $40$.
These modes lie at wave numbers comparable to the cutoff of the grid and extend across the domain, with $\langle x^2\rangle\approx290$ at $t=0$ on $|x|\le32$, and we report only multipliers whose eigenvectors satisfy $\langle x^2\rangle<150$ at $t=0$.
This retains the modes of the cloud, whose eigenvectors have $\langle x^2\rangle<100$, together with the lower trap levels, and the largest multipliers reported do not change when the threshold is set anywhere between $60$ and $150$.

For $g=100$, $p_0=10$ the multipliers of the cloud lie on the unit circle to within $10^{-3}$ in the spectra computed at seven gains between $\gamma=0.029$ and $0.03826$.
At $\gamma=0.03835$ a complex quadruplet has left the circle, and from $\gamma=0.0384$ on also a real pair through $-1$.
At the turn of the $(50,10)$ family the largest multiplier belongs to a complex quadruplet, of modulus $1.05$.
At the turn of the $(35,10)$ family the largest quadruplet has modulus $1.01$, comparable to the largest modulus among the multipliers of the domain in the same spectrum, $1.009$.
The spectrum at the turn of the $(100,3)$ family is computed with the gain tapered between $|x|=24$ and $34$, where this orbit holds no atoms, and all its multipliers of the cloud lie on the unit circle.

Near $+1$ the pair that approaches it is distinguished from the four multipliers held at $+1$, which lie within $2\times10^{-3}$ of $+1$ at the filled points of Fig.~\ref{fig:fold}(b), and from other modes of the cloud whose phase over a period is close to a multiple of $2\pi$, by the overlap of the plane spanned by the real and imaginary parts of its eigenvectors with the tangent of the family, estimated from the difference of neighboring orbits.
Below the resonance this overlap lies between $0.72$ and $0.97$ for this pair and below $0.45$ for the other pairs on the unit circle within $0.4$ of $+1$ in phase.
On $|x|\le32$ with the tapered gain the pair stays at $e^{\pm0.04i}$ up to $\gamma=0.03853$, the square in Fig.~\ref{fig:fold}(b).

Direct time evolution confirms these spectra.
A random perturbation of relative norm $10^{-6}$, smooth on the scale of the trap length, is added at $t=0$ to an orbit whose closure after one period has been brought to $10^{-10}$ in relative norm, and the distance $\|e^{-i\vartheta}\psi-\psi_0\|/\|\psi_0\|$ of the evolved state from the orbit is recorded once per period, minimized over the global phase $\vartheta$ and, once the distance saturates, also over a shift in time along the orbit.
At the turn on $|x|\le40$ it grows by the factor $1.147$ per period between the $40$th and the $80$th period, against $1.148$ for the largest multiplier.

\subsection{Discretization}

The evolution is computed with a second-order split-step Fourier scheme on the domain $|x|\le32$, with $N_x=1025$ grid points, spacing $\Delta x=0.0625$, and $4000$ time steps per period; near the critical gain we use the domain $|x|\le40$ with $N_x=1281$, at the same spacing.
The kinetic step is exact in Fourier space, and the potential step is the exact solution of the local equation $i\partial_t\psi=(\tfrac12x^2+i\gamma x+g|\psi|^2)\psi$ over a half step $\tau$, in which the density grows as $e^{2\gamma x\tau}$ and the nonlinear phase is $g|\psi|^2(e^{2\gamma x\tau}-1)/(2\gamma x)$.
This choice is essential, because the usual exponential of the potential evaluated at the initial density is not exactly reversible in the presence of gain, its nonlinear phase being in error by $g|\psi|^2\gamma x\tau^2$ per step, a fraction $\gamma x\tau$ of the phase increment, and the half-period shooting of Appendix~\ref{app:numerics}\,1 then fails to close the orbit.
With the exact local step the scheme is reversible to round-off in each step, and $\mathcal U_{-T}\mathcal U_T$ differs from the identity by $10^{-10}$.

Two constraints set the resolution.
The split-step scheme has a nonlinear instability of the highest wave numbers when $k_{\max}^2\Delta t/2>\pi$~\cite{weideman1986split}, where $k_{\max}=\pi/\Delta x$ is the largest wave number of the grid.
The linearized dynamics used for the Jacobian and the monodromy amplifies this instability from round-off, so we keep $\Delta t<2\Delta x^2/\pi$ throughout.
The spectrum of the nonlinear term, which extends to about three times the largest wave number carried by $\psi$, has to stay well inside the grid cutoff, and this bounds the range of $g$ and $p_0$ that a given grid can treat; the six families of Sec.~\ref{sec:foldsec} lie within it at $\Delta x=0.0625$.

Grid convergence was verified by repeating orbits of the $g=100$, $p_0=10$ family with $N_x=2049$ and $16000$ steps; the periods agree with those of the standard grid to $2\times10^{-7}$ in $T/2\pi$, half of which is the shift of the Kohn period by the time step given below; closer to the critical gain the position of the turn depends on the domain through the resonances of Appendix~\ref{app:numerics}\,2.
The kicked cloud of Sec.~\ref{sec:experiment} is evolved on $|x|\le48$ with $\Delta x=0.04$ and $8000$ steps per trap period, with the gain tapered to zero between $|x|=32$ and $44$ as in Appendix~\ref{app:numerics}\,2.

The time step shifts the period of the Kohn orbit itself, by the relative amount $-(\pi/4000)^2/6=-1.0\times10^{-7}$, which is $0.6\%$ of the shift at $\gamma=0.001$; $c_2$ is therefore evaluated with $T$ measured against that period, and its limit $\gamma\to0$ in Table~\ref{tab:coef} is obtained from a quadratic fit in $\gamma$ to the first four orbits of each family, linear and cubic fits giving the same limit to within $0.2\%$.

The five-variable model of Eq.~\eqref{eq:model} inherits the reversibility of Eq.~\eqref{eq:gpe}, being invariant under $(t,X,P,S,C,N)\to(-t,-X,P,S,-C,N)$, so that its symmetric periodic orbits cross $X=C=0$ twice per period.
With $N=1$ and $\dot X=p_0$ at the crossing, such an orbit is determined by $S_0$ and $T$, which a two-dimensional Newton iteration on the half-period map finds; the map is integrated by fourth-order Runge--Kutta with $600$ steps, and the family is continued in $\gamma$ with the step halved whenever the iteration fails, which locates the fold, and is followed past it by pseudo-arclength continuation in $(S_0,T,\gamma)$.
Doubling the number of steps changes the period by less than $10^{-8}$ and moves the fold of the model by less than the last digit quoted in Table~\ref{tab:fold}.

\section{Scaling of the critical gain}\label{app:fits}

The exponents $\nu_A$ and $\nu_R$ of Sec.~\ref{sec:scaling} come from a least-squares fit of $\ln\gamma_{\rm c}$ to $\ln A$ and $\ln R_{\rm TF}$ over the six families of Table~\ref{tab:fold}, which leaves three degrees of freedom, and their uncertainties are the standard errors of this fit.
If $\gamma_{\rm c}R_{\rm TF}$ were a power of $p_0/R_{\rm TF}$, then with $A\simeq p_0$, which holds to $20\%$ in Table~\ref{tab:fold}, the exponents would satisfy $\nu_A+\nu_R=1$, whereas the fit gives $\nu_A+\nu_R=0.73(9)$.
With this constraint imposed the fit gives $\nu_A=0.51(3)$, and the six critical gains scatter about it by $3.0\%$ rms, as they do by $3.1\%$ about Eq.~\eqref{eq:foldlaw}.
The departure from the constraint is carried by the three families at $p_0=10$, for which $\gamma_{\rm c}\sqrt{AR_{\rm TF}}$ decreases from $0.293$ at $g=100$ to $0.265$ at $g=35$.
Taking $\gamma_{\rm c}$ of each family instead from the largest gain reached on any of the domains of Appendix~\ref{app:numerics}\,2 changes the mean value $0.280$ in Eq.~\eqref{eq:foldlaw} by $0.4\%$ and $\nu_R$ from $0.27$ to $0.28$.

In the $(\varepsilon_1,\varepsilon_2)$ plane of Fig.~\ref{fig:fold}(d) the fifty folds of the model lie within $1.3\%$ in $\varepsilon_2$ of a smooth curve whose logarithmic slope steepens with $\varepsilon_1$.
The critical gains of the six GP families scatter about the power law fitted to them, of slope $-2.7(2)$, by $11\%$ rms in $\varepsilon_2$; they cover only a factor $2.2$ in $\varepsilon_1$, too little to determine the shape of the curve that they trace.

\input{main.bbl}
\end{document}

%% file: main.bbl
%